\documentclass [aps,showpacs,showkeys,preprint,amsmath,amssymb]{revtex4}
\usepackage{graphicx}
\usepackage{amsmath}

\begin{document}

\title{ \bf Universal quantum computation with electron spins in quantum dots based on superpositions of space-time paths and Coulomb blockade}

\author{\bf Cyrus C.~Y.~Lin$^{a}$, Chopin Soo$^{a}$, Yin-Zhong Wu$^{a,~b}$, and Wei-Min Zhang$^{a}$}

\affiliation{$^{a}$Department of Physics and Center for Quantum
Information Science, \\ National Cheng Kung
University, Tainan 70101, Taiwan\footnote{Email: yzwu@phys.ncku.edu.tw}\\
 $^{b}$Department of Physics, Changshu Institute of Technology,
Changshu 215500, China}

\begin{abstract}

Using electrostatic gates to control the electron positions, we
present a new controlled-NOT gate based on quantum dots. The qubit
states are chosen to be the spin states of an excess conductor
electron in the quantum dot; and the main ingredients of our
scheme are the superpositions of space-time paths of electrons and
the effect of Coulomb blockade. All operations are performed only
on individual quantum dots and are based on fundamental
interactions. Without resorting to spin-spin terms or other
assumed interactions, the scheme can be realized with a dedicated
circuit and a necessary number of quantum dots. Gate fidelity of
the quantum computation is also presented.
\end{abstract}


\maketitle
\section{Introduction}
The study of quantum computing and quantum information processing
has received great attentions in the past decade\cite{book1}.
Among several proposals of implementing quantum information
processing in realistic physical systems, solid state quantum
computing is one of the most promising candidates for future
quantum computer architecture due to its scalability. Meanwhile,
the idea of using spin degree of freedom in
electronics\cite{book2} has been strongly supported from recent
experiments showing unusually long spin decoherence time in
semiconductors. The first QC scheme based on electron spin in
semiconductor QD was proposed by Loss and
Divincenzo\cite{PRA1998}, in which the controlling operations are
effected by gating of the tunnelling barrier between neighboring
dots, and one-qubit operations by applied magnetic field. Kane
designed a silicon based nuclear spin quantum computer\cite{Kane},
information is encoded into the nuclear spins of donor atoms in
doped silicon electronic devices. Local operations on individual
spins are performed using externally applied electric fields, and
interaction between qubits is mediated through the donor-electron
exchange interaction. In Ref.~\cite{Sham}, optical RKKY
interaction between charged semiconductor QDs is generated via
virtual excitation of delocalized excitons, and it provides an
efficient coherent control of the spins. Very recently, Calarco et
al.\cite{PRA2003} presented an all optical implementations of QC
with semiconductor QDs, where two-qubit gates are realized by
switching on trion-trion coulomb interactions between different
QDs, state selectivity is achieved via conditional laser
excitation exploring Pauli exclusion principle.

The idea of most existing QC schemes is to design an effective
Hamiltonian and then find or adjust a physical system to evolve
approximately under the effective Hamiltonian. Any deviation from
the "beautiful" effective Hamiltonian in practical physics system
will bring complexity, even disaster, to the practical
QC\cite{XHu}. In this paper, we give a new QC scheme based on a
more fundamental Hamiltonian in solid state physics. The qubit
states are selected as the spin states of an excess electron
stored in a semiconductor QD. The Hamiltonian of electrons in
semiconductor QDs dressed by an external magnetic field can be
described as following
\begin{eqnarray}
H&=&\sum_{i}h_{i},\\\nonumber
h_{i}&=&\frac{1}{2m_{e}^{*}}P_{i}^{2}-g\mu_{B}{\bf B_{i}\cdot
S_{i}}+V(r_{i}).
\end{eqnarray}
where single-particle Hamiltonian $h_{i}$ describes the electron
dynamics confined to semiconductor QD. $m_{e}^{*}$ is the
effective mass of electron in semiconductor QD. We allow for a
local magnetic field $\bf{B}_i$ applied on QD $i$. $V(r_{i})$ is
the confinement-potential experienced by the electron in QD.

All operations in our scheme are performed on individual QDs.
Using single-qubit rotation and adjusting gate voltage to drive
electron between QDs through one dimensional quantum wire, a
quantum control gate can be realized naturally. The detailed
descriptions on design of a CNOT gate are given in Sec.~II.
 Physical implementation and a brief discussion on gate fidelity are presented
 in Sec.~III. Our conclusions are summarized in Sec.~IV.

\section{Formula for quantum gate}
It is well known that single qubit and CNOT gates together can be
used to implement an arbitrary unitary operation, and therefore
are universal for quantum computation\cite{book1}. A single qubit
is a vector $|\psi\rangle=a|0\rangle+b|1\rangle$ parameterized by
two complex numbers satisfying $|a|^{2}+|b|^{2}=1$. Operations on
a qubit must preserve this norm, and thus are described by
$2\times 2$ unitary matrices. Any single qubit rotations can be
realized by two of the following three rotation operators
$R_{x}(\theta)=e^{-i\theta \sigma_{x}}$,
$R_{y}(\theta)=e^{-i\theta \sigma_{y}}$, and
$R_{z}(\theta)=e^{-i\theta \sigma_{z}}$, where $\sigma_{x}$,
$\sigma_{y}$, and $\sigma_{z}$ are Pauli matrices.

In textbook, a CNOT operator is described by a $4\times 4$ matrix,
\begin{equation}
CNOT=\left(%
\begin{array}{cccc}
  1& 0 & 0 & 0 \\
  0& 1 & 0 & 0 \\
  0& 0 & 0 & 1 \\
  0& 0 & 1 & 0 \\
\end{array}%
\right),
\end{equation}
it can be rewritten as

\begin{equation}
CNOT=\frac{1}{2}[I\otimes\sigma_{x}-\sigma_{z}\otimes\sigma_{x}+I\otimes
I +\sigma_{z}\otimes I],
\end{equation}
where $I$ is $2\times 2$ identity matrix. Further, substituting
Eq.~(3) into
 $CNOT=-ie^{i\frac{\pi}{2}CNOT}$, we have
\begin{eqnarray}
CNOT&=&e^{-i\frac{\pi}{2}I\otimes I}\cdot
e^{i\frac{\pi}{4}(I\otimes\sigma_{x}-\sigma_{z}\otimes\sigma_{x}+I\otimes
I +\sigma_{z}\otimes I)},\nonumber\\
 &=&e^{-i\frac{\pi}{4}I\otimes I}\cdot e^{i\frac{\pi}{4}I\otimes \sigma_{x}}\cdot e^{i\frac{\pi}{4}\sigma_{z}\otimes I}\cdot e^{-i\frac{\pi}{4}\sigma_{z}\otimes
 \sigma_{x}},\nonumber\\
 &=&e^{-i\frac{\pi}{4}I\otimes I}\cdot e^{i\frac{\pi}{4}I\otimes \sigma_{x}}\cdot e^{i\frac{\pi}{4}\sigma_{z}\otimes I}\cdot\frac{1}{\sqrt{2}}(I\otimes I-i\sigma_{z}\otimes
 \sigma_{x}),\\
&=&e^{-i\frac{\pi}{4}I\otimes I}\cdot e^{i\frac{\pi}{4}I\otimes
\sigma_{x}}\cdot e^{i\frac{\pi}{4}\sigma_{z}\otimes
I}\cdot\frac{1}{\sqrt{2}}[I\otimes I-i(\sigma_{z}\otimes
I)\cdot (I\otimes\sigma_{x})],\nonumber\\
&=&e^{-i\frac{\pi}{4}I\otimes I}\cdot e^{i\frac{\pi}{4}I\otimes
\sigma_{x}}\cdot e^{i\frac{\pi}{4}\sigma_{z}\otimes
I}\cdot\frac{1}{\sqrt{2}}(I\otimes I+e^{-i\frac{\pi}{2}}\cdot
e^{-i\frac{3\pi}{2}\sigma_{z}\otimes I}\cdot
e^{-i\frac{\pi}{2}I\otimes\sigma_{x}}).\nonumber
\end{eqnarray}

The first term at the last line of Eq.~(4) can easily be realized
by letting two qubits evolve freely under each single electron
Hamiltonian respectively without any operations. The second and
the third terms are only single spin rotations. By the
superposition of electron spin state and charge position, the
fourth term in Eq.~(4) can be implemented naturally, which will be
illustrated in Sec.~III. So, only six steps(including five
operations) are required to complement a CNOT quantum gate given
by Eq.~(4). Compared with the CNOT gate presented in
ref.~\cite{PRA1998}, there is no need for spin swap operation in
our scheme, and operations are performed on individual QDs here.
Furthermore, four terms on the right side of Eq.~(4) commute each
other, we can
 exchange the operator sequence arbitrarily.

\section{Physical implementation}
\begin{figure}[ht]
   \centering
   \includegraphics[width=3in]{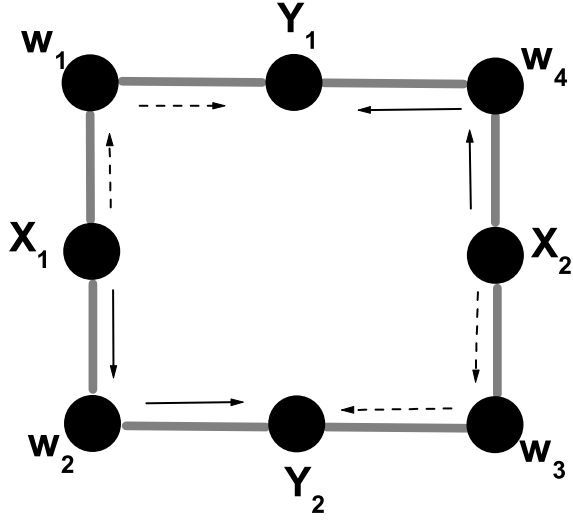}
   \caption{Architecture for implementing a CNOT gate. Black spheres stand for quantum dots,
   and gray lines stand for one dimensional quantum wires.}
\end{figure}
The spin states of the excess conductor electron of QD are
employed as qubits. Manipulating a single spin can be done by
applying a pulsed local magnetic field, two-qubit control gate in
this paper is realized by a specific spacial distribution of QDs.
The architecture of our designed CNOT gate is shown in Fig.~1,
where the gray thick lines stand for one-dimension quantum wires,
which are used to coherently transport electrons between
QDs\cite{Flying}. Eight QDs are needed to implement a CNOT gate.
Initial states of two qubits(control qubit and target qubit) are
stored in spin states of the excess electron in QD $X_1$ and $X_2$
respectively, and final states of the two qubits are defined as
the spin states of electron in QDs $Y_1$and $Y_2$. QDs $w_1$,
$w_2$, $w_3$, $w_4$, $Y_1$, and $Y_2$ are all empty of electrons
in the initial time. Firstly, the electron is being driven from QD
$X_1$ to QD $w_1$ and $w_2$ by adjusting gate voltage. The design
of electric gates can refer to the principle of single electron
tunnelling(SET) transistor\cite{SET} and the flying qubit in
one-dimensional quantum wires\cite{Flying}. Assuming the two paths
and two QDs ($w_1$ and $w_2$)are symmetric, the probabilities of
electron in QD $w_1$ and $w_2$ will be equal in theory. By the
same method, The electron in QD $X_2$ is also being driven into QD
$w_3$ and $w_4$ in parallel. Secondly, we apply an external local
magnetic field along z direction on QD $w_1$ to do a spin rotation
$e^{-i\frac{3\pi}{2}\sigma_{z}}$. Thirdly, an external local
magnetic field along $x$ direction on QD $w_4$ is used to
complement operator $e^{-i\frac{\pi}{2}\sigma_{x}}$. Fourthly,
apply two electric fields on QD $w_1$ and QD $w_4$ respectively to
create a global phase difference $e^{-i\frac{\pi}{2}}$ between QDs
$w_1$, $w_4$ and QDs $w_2$, $w_3$. At last, two electrons are
pushed from QDs $w_1$, $w_4$ to QD $Y_1$, and from QDs $w_2$ and
$w_3$ to QD $Y_2$ respectively. Adjusting gate voltage to
eliminate the case that two electrons site on QD $Y_1$ or $Y_2$
simultaneously. According to the coulomb blockade effect, there
are only two possible pathes for the two electrons actually occur.
One is $X_1\rightarrow w_2\rightarrow Y_2$; $X_2\rightarrow
w_4\rightarrow Y_1$(Solid arrows in Fig.~1), The other is
$X_1\rightarrow w_1\rightarrow Y_1$; $X_2\rightarrow
w_3\rightarrow Y_2$(Dashed arrows). In mathematic language, the
above process can be described as following,
\begin{equation}
|S_{X_1}\rangle\otimes|S_{X_2}\rangle=[\rho_{w_1}e^{-i\frac{\pi}{2}}e^{-i\frac{3\pi}{2}\sigma_{z}}|S_{X_1}\rangle
+\rho_{w_2}|S_{X_1}\rangle]\bigotimes
[\rho_{w_3}e^{-i\frac{\pi}{2}}e^{-i\frac{\pi}{2}\sigma_{x}}|S_{X_2}\rangle+\rho_{w_4}|S_{X_2}\rangle],
\end{equation}
where $\rho_{w_1}$ and $\rho_{w_2}$ are the probabilities of
electron in QD $w_1$ and QD $w_2$ respectively,
 $\rho_{w_3}$ and $\rho_{w_4}$ are the probabilities of electron in QD $w_3$ and QD $w_4$ respectively. Further,
 we expand Eq.~(5) by summation of possible pathes mentioned
above, only two terms are retained,
\begin{eqnarray}
|S_{X_1}\rangle\otimes
|S_{X_2}\rangle&=&[\rho_{w_1}\rho_{w_3}e^{i\frac{\pi}{2}}e^{-i\frac{3\pi}{2}\sigma_{z}\otimes
I}\cdot
e^{-i\frac{\pi}{2}I\otimes\sigma_{x}}|S_{X_1}\rangle\otimes
|S_{X_2}\rangle+\rho_{w_2}\rho_{w_4}I\bigotimes
I|S_{X_1}\rangle\otimes|S_{X_2}\rangle],\nonumber\\
&=&[\rho_{w_1}\rho_{w_3}e^{i\frac{\pi}{2}}e^{-i\frac{3\pi}{2}\sigma_{z}\otimes
I}\cdot
e^{-i\frac{\pi}{2}I\otimes\sigma_{x}}+\rho_{w_2}\rho_{w_4}I\bigotimes
I]|S_{X_1}\rangle\otimes|S_{X_2}\rangle.
\end{eqnarray}

We can see that the fourth term in Eq.~(4) can be complemented by
the above architecture under the condition of equal
probabilities($\rho_{w_1}=\rho_{w_2}=\rho_{w_3}=\rho_{w_4}$). Two
single-qubit operations (the 2nd and 3rd terms in Eq.~(4))can be
performed by applied two local pulsed magnetic fields along $X$
and $Z$ directions on QDs $X_1$ and $X_2$ respectively at the
initial time. They can also be performed on QDs $Y_1$ and $Y_2$ at
the final time. The successful criterium of the above control gate
is that two electrons must arrive at QDs $Y_1$ and $Y_2$ at the
final time. If QDs $Y_1$ and $Y_2$ only process one electron at
the final time($50\%$ probability), then the control operation
will be discarded. So, a measurement must be done at the final
time to check whether the control gate has been performed
successfully.

If a two-qubit control gate has been done successfully, the
potential quantum error of our scheme comes from the asymmetric
probabilities of electron positions, namely,
$\rho_{w_1}\neq\rho_{w_2}$, $\rho_{w_3}\neq\rho_{w_4}$. We define
the fidelity of our CNOT gate induced by the asymmetric
probabilities of electron positions as

\begin{equation}
F=\{
\begin{array}{cc}
   2-U^{\dagger}U^{'}, & U^{\dagger}U^{'}\geq 1; \\
  U^{\dagger}U^{'}, & U^{\dagger}U^{'}< 1, \\
\end{array}
\end{equation}
and
\begin{eqnarray}
U^{\dagger}U^{'}&=&{\frac{\rho_{w_1}\rho_{w_3}\rho_{w_1}'(x)\rho_{w_3}'(x')
+\rho_{w_2}\rho_{w_4}\rho_{w_2}'(1-x)\rho_{w_4}'(1-x')}{(\rho_{w_1}\rho_{w_3})^{2}
+(\rho_{w_2}\rho_{w_4})^{2}}}, \\
\rho'(x)&=&x,
\end{eqnarray}
where operator $U$ is the ideal CNOT operator described by
Eq.~(6), $U'$ is the practical operator with the consideration of
potential asymmetric probabilities of electron positions in real
space. under the case of equal probabilities (symmetric case), we
have $\rho^{'}_{w_i}=\rho_{w_i}=0.5$, and the fidelity is equal to
$1$ by the definition of Eq.~(7). we use $x (0\leq x\leq 1)$to
measure the magnitude of asymmetry of electron positions, and the
symmetric case corresponds to $x=0.5$. Fidelity of the CNOT gate
as a function of the asymmetry in case $x=x'$ is plotted in
Fig.~2. It is shown, in Fig.~2, that fidelity will be up $99\%$
when the asymmetry does not exceed 10\%.
\begin{figure}[ht]
   \centering
   \includegraphics[width=3in]{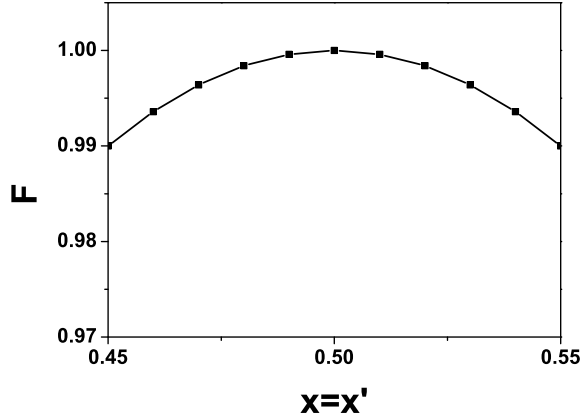}
   \caption{The fidelity of the CNOT gate as a function of the asymmetry of electron positions}
\end{figure}

Under the precondition of high fidelity, a small size difference
between QD $w_1$ and $w_2$( and between QD $w_3$ and $w_4$) is
permitted. Reminding that QDs $w_1$ and $w_4$ does not need the
same in size. As a result, the single-qubit operations can be
performed selectively with different frequencies, and the local
pulsed magnetic field may also be replaced by optical operations,
and the single qubit rotations can be proceeded by applying two
laser fields which exactly satisfy the Raman-resonance condition
between spin-up state and spin-down
state\cite{adiabatic}\cite{PRL1}. Optical
operation will sufficiently reduce single-qubit operation time.\\

Operation time and decoherence time are critical criteria on the
feasibility of a QC scheme\cite{book1}. For a single spin flip
rotation, a switching time of $\tau_{s}=30ps$ is required, using a
magnetic field of $B=1T$\cite{SSC}. Considering the switching
on/off time of electric field is about $20\sim 30ps$, gate voltage
must be adjusted carefully to create a small potential difference
between QDs $w_{1}, w_{4}$, and QDs $w_2, w_{3}$\cite{Poisson}. If
two pulsed lasers are used to complement a single qubit operation,
the single spin flip operation time is about ten picoseconds for a
given coupling constant between Laser and QD $\Omega=1meV$.
Decoherence time of spin states in semiconductor QD has also been
extensively studied in theory and experiment\cite{GaAs}\cite{GaN},
and coherent transport of spin charge in semiconductor has been
discussed in detail.~\cite{book2}\cite{Flying}\cite{SSC}. State
initialization and read-out can be implemented using spin filters
and polarizers in any designs\cite{SSC}\cite{RI}. Our motivation
is to show a new theoretical design direction on quantum control
gate based on semiconductor QDs. So, we do not focus more
attentions on challenges common to all other implementations in
electron spins and QDs system. The great challenge to our scheme,
which is different from other schemes in practical, is how to
manipulate single electron transport coherently between QDs
through one-dimensional quantum wire or nanotube, this difficulty
can be solved by the rapid development of single electron
tunnelling transistor technologies.
\section{Concluding Remarks}

In summary, we have presented a new QC scheme based on
semiconductor QDs system. The superposition of spin state and
charge position in real space and coulomb blockade effect are
introduced to design a CNOT gate. All the quantum states'
evolutions in our scheme base on a more fundamental Hamiltonian.
No direct spin-spin coupling is needed here. Our scheme, which is
based on solid state QD, is essentially different from the mobile
qubit scheme\cite{Flying}, and spin state operations are more
stable in general. As we know, it is the first time that a control
gate is designed by the superposition of electron spin state and
charge position in quantum computation based on QDs system. Our
scheme is simple except for a dedicated circuit and more QDs
needed. Furthermore, our scheme can safely be extended into the
charge qubit case by replacing each QD with a coupled QDs, and
qubit states are defined as the excess electron charge states in
coupled QDs.
\begin{acknowledgments}
This work was partially supported by the National Science Council
of ROC under Contract No.~NSC-93-2120-M-006-005,
No.~NSC-93-2112-M-006-011, No.~NSC-93-2112-M-006-019 and National
Center for Theoretical Science(South), Taiwan.
\end{acknowledgments}

\end{document}